\pdfoutput=1

\documentclass[11pt]{article}

\usepackage[]{ACL2023}

\usepackage{times}
\usepackage{latexsym}

\usepackage[T1]{fontenc}

\usepackage[utf8]{inputenc}

\usepackage{microtype}

\usepackage{inconsolata}

\usepackage{hyperref}
\usepackage{url}
\usepackage{graphicx}
\usepackage{adjustbox}
\usepackage{booktabs} 
\usepackage{multirow}
\usepackage{amsmath,mathtools}
\usepackage{amssymb}
\usepackage{amsfonts}
\usepackage{subfigure}
\usepackage{caption}
\usepackage{colortbl}
\usepackage{changepage}
\usepackage{xspace}
\usepackage{array}
\newcommand{\our}{GeAR}
\newcommand{\logo}{\raisebox{-1pt}{\includegraphics[width=1.7em]{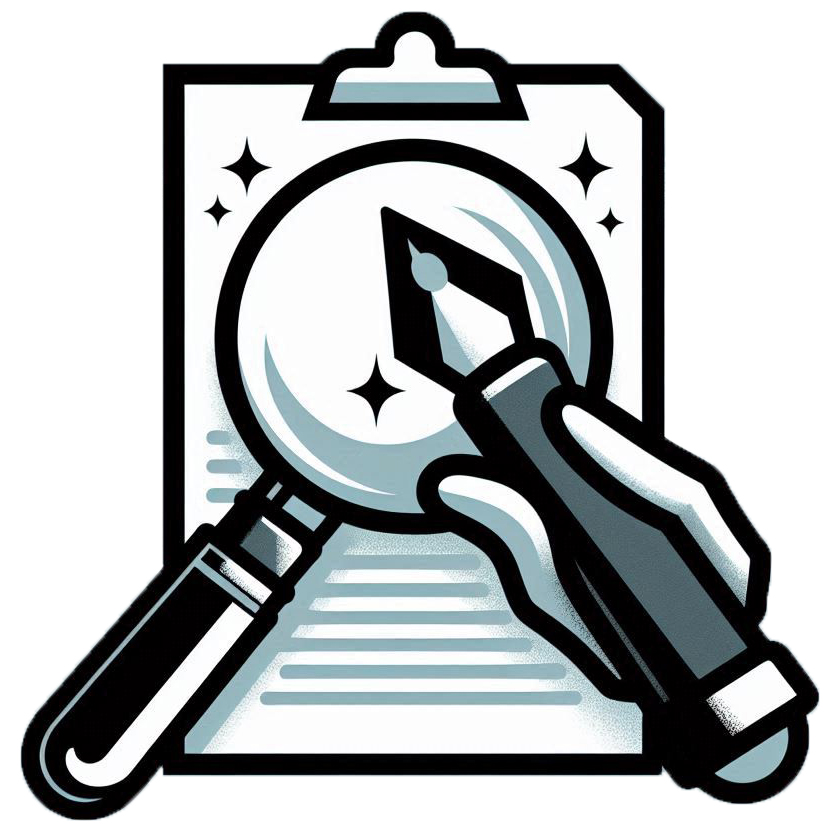}}\xspace}

\title{\logo~\our: Generation Augmented Retrieval}

\author{\bf Haoyu Liu, Shaohan Huang, Jianfeng Liu, Yuefeng Zhan, \\
{\bf Hao Sun, Weiwei Deng, Feng Sun, Furu Wei, Qi Zhang}\\
  Microsoft Corporation \\
  \small{implhy@gmail.com} \\
  \small{\{shaohanh, jianfengliu, yuefzh, hasun, dedeng, sunfeng, fuwei, qizhang\}@microsoft.com} \\
}

\begin{document}
\maketitle
\begin{abstract}
Document retrieval techniques are essential for developing large-scale information systems. The common approach involves using a bi-encoder to compute the semantic similarity between a query and documents. However, the scalar similarity often fail to reflect enough information, hindering the interpretation of retrieval results. In addition, this process primarily focuses on global semantics, overlooking the finer-grained semantic relationships between the query and the document's content. In this paper, we introduce a novel method, $\textbf{Ge}$neration $\textbf{A}$ugmented $\textbf{R}$etrieval ($\textbf{GeAR}$), which not only improves the global document-query similarity through contrastive learning, but also integrates well-designed fusion and decoding modules. This enables GeAR to generate relevant context within the documents based on a given query, facilitating learning to retrieve local fine-grained information.
Furthermore, when used as a retriever, GeAR does not incur any additional computational cost over bi-encoders. GeAR exhibits competitive retrieval performance across diverse scenarios and tasks. Moreover, qualitative analysis and the results generated by GeAR provide novel insights into the interpretation of retrieval results. The code, data, and models will be released at \href{https://github.com/microsoft/LMOps}{https://github.com/microsoft/LMOps}.
\end{abstract}

\section{Introduction}

Document retrieval serve as the foundational technology behind large-scale information systems, playing a crucial role in applications such as web search, open-domain question answering (QA)~\citep{drqa, dpr}, and retrieval-augmented generation (RAG)~\citep{rag, liu-etal-2024-se2, rag_survey}. The predominant approach in passage retrieval is to construct a bi-encoder model~\citep{sbert}. In this framework, queries and documents are encoded separately, converting each into vector representations that enable computation of their semantic similarity in a high-dimensional space. 


However, this similarity calculation process faces several challenges. 
First, the complex semantic relationship between query and document is mapped to a scalar similarity, which cannot reflect enough information and is difficult to understand~\cite{maxsime}.
Second, when dealing with long documents, such as those with 256, 512, or even more tokens, identifying the section most relevant to the query and contributing most to the similarity is highly desirable but challenging to achieve~\cite{bge_landmark, late_chunking}.
Moreover, many NLP tasks, such as sentence selection, search result highlighting, needle in a haystack~\citep{lost, film, leave}, and fine-grained citations~\citep{cite, longcite}, require a deep and fine-grained understanding of the text. Given this need for fine-grained understanding, the bi-encoder that simply aligns the full document to the query seems insufficient, as its conventional contrastive loss mainly emphasizes global semantics~\cite{colbert}. 
To complement this core capability of the retriever, we propose a novel and challenging fundamental question: \textit{How to make the retriever have both \textbf{global} and \textbf{local} understanding and retrieval capabilities?}

Although the concept is intutive, several challenges remain. First, it is difficult to construct sufficient data to support effective solutions to this problem in previous research work. Second, the training objectives, model architectures, design details, as well as how to effectively train the models, have not been fully explored. To address these challenges, we propose a novel approach \textbf{\our} (\textbf{Ge}neration-\textbf{A}ugmented \textbf{R}etrieval). In it, we build a pipeline to efficiently synthesize large amounts of high-quality (query-document-information) triples by utilizing large language models. 
In terms of method, \our~retains to leverage contrastive learning to optimize the similarity between the query and the global document. 
To improve the interaction between local information and queries, we design a text decoder that generates fine-grained information from the document in response to a given query. This enhances the model’s ability to understand local semantics. In this way, \our~can handle both the retrieval of global documents and local information simultaneously.

We conduct extensive experiments on two retrieval tasks, and compared with the BGE and BGE-Reranker-L, \our~achieves 3.5\% and 12.9\% relative improvements on global document retrieval and local information retrieval tasks respectively. \our's versatility and visual analysis also shed new light on the interpretability and comprehensibility of retrieval results.


Overall, our contributions are summarized as follows:

\begin{itemize}
    \item We introduce a new global-local retrieval task, which presents challenges for both document retrieval and fine-grained information retrieval within documents.
    
    
    \item We introduce \our, 
    which augmented the model's global and local understanding and retrieval capabilities of documents by incorporating a generation task. 
    
    \item Through extensive experiments, GeAR has shown competitive performance across various retrieval tasks. GeAR's versatility also makes the retrieval results more explainable.
\end{itemize}

\begin{figure*}[ht]
\begin{center}
  \adjustbox{margin=-0.2cm 0cm 0cm 0cm}{
    \includegraphics[scale=0.48]{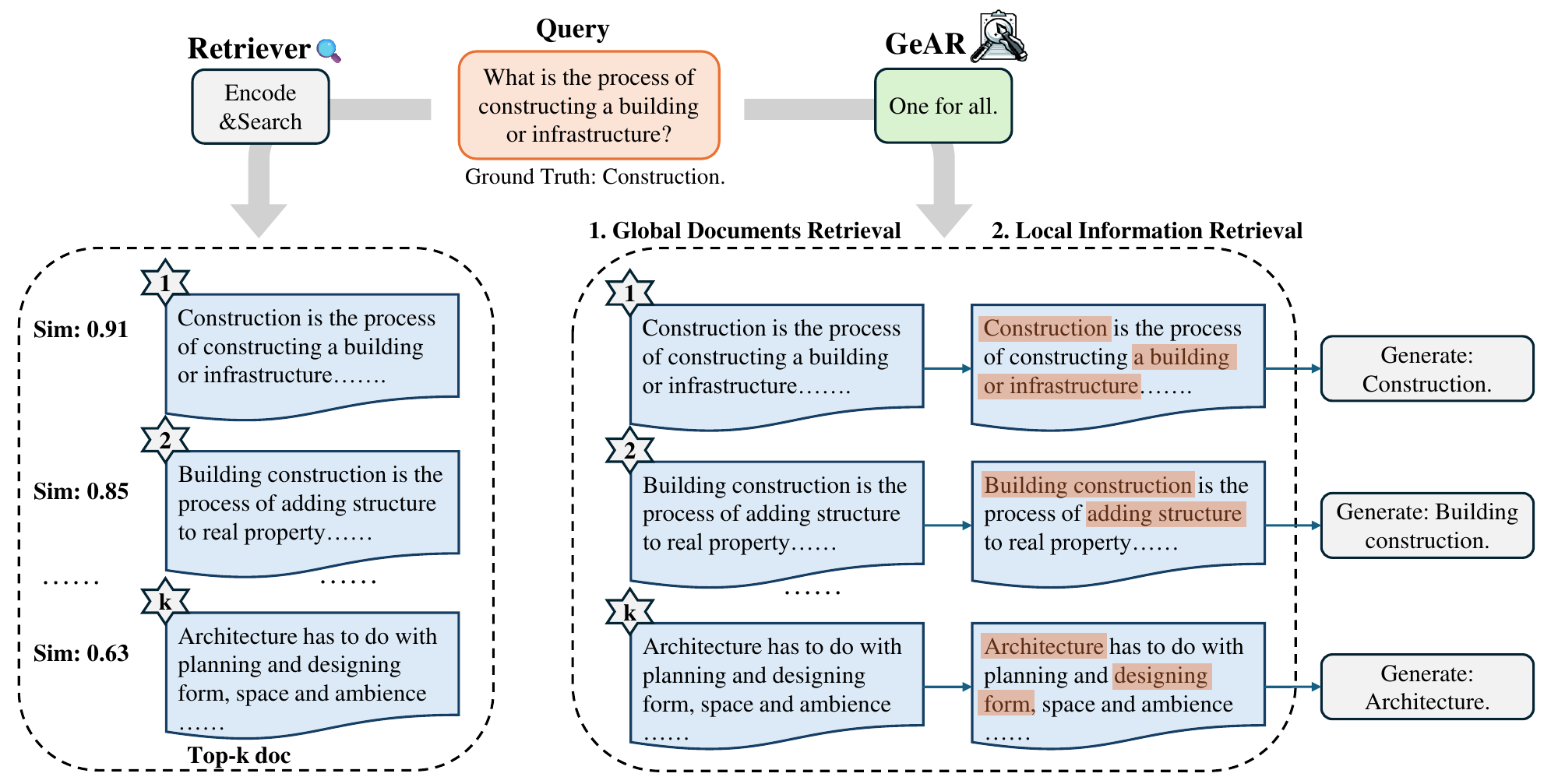}
  }
  \caption{Comparison of functionality between classical retriever and \our. \our~is designed to handle both global document retrieval and local information retrieval simultaneously. In addition, \our~can generate information based on the query for reference.}
  \label{fig:\our}
\end{center}
\end{figure*}

\section{Related Work}

\subsection{Embedding-based Retrieval}
Embedding-based retrieval has emerged as a cornerstone of modern information retrieval systems, enabling efficient semantic search through dense vector representations. Early approaches like Word2Vec~\citep{word2vec} and GloVe~\citep{glove} demonstrated the potential of learning distributed word representations, while more recent transformer-based models such as BERT~\citep{devlin-etal-2019-bert} have pushed the boundaries of contextual embeddings. Bi-encoder architectures~\citep{sbert} have become particularly popular for retrieval tasks~\cite{dssm}. Recent advances include contrastive learning objectives~\citep{dpr, e5, gte, simcse} and hard negative mining strategies~\citep{xiong2021approximate} to improve embedding quality. \citet{grit} explored how to generate text and provide excellent semantic representation by distinguishing task instructions. 
Multimodal information retrieval also relies on high-quality semantic representations, where the embedding space serves to bridge different modalities, including text, images, and video. Vision language models such as CLIP~\citep{clip}, ALBEF~\citep{albef}, and BLIP~\citep{li2022blip} have demonstrated remarkable zero-shot capabilities by learning joint embeddings derived from large scale image-text pairs. 

\subsection{Fine-grained Information Mining}
Mining fine-grained information in a long context during retrieval has become a key challenge for efficient information retrieval. The naive heuristic hierarchical approach involves further chunking documents and then calculate semantic similarity with the query on the chunked sentences. However, finer chunking easily leads to increased computational complexity and semantic incoherence~\cite{HAN, dense_hierarchical, hybrid}. 
In question-answering tasks, RNN or BERT is often used to compute token representations and train classifiers for information extraction~\citep{baf, match-lstm, drqa, bert-rc}. 
With the development of generative models, there have been many efforts to enhance the model's ability to find a needle in a haystack~\citep{lost, film, leave}. Another similar task is to have the model add reference information to the original text when generating responses~\citep{cite, longcite}. Coincidentally, some recent research is dedicated to improving the region-level understanding ability of multimodal large language models (MLLMs)~\citep{loc-clip}. 

Despite these advances, we find that these works often rely on heavy decoder-only models that are independent of the retrieval model, but few focus on mining fine-grained information during the retrieval stage.

\section{Generation Augmented Retrieval}


\subsection{Preliminaries}
In this work, we formalize the global-local retrieval task as follows: Let a document corpus as $\mathbb{D}$, which contains $N$ documents $\{d_1,...,d_i,...,d_N\}$. Each of these documents $d_i$ contains a number of fine-grained information units $\{u_{1},...,u_{l_i}\}$, such as sentences, where $l_i$ is the units number of $d_i$. 
Our goal is to find a retrieval method $f(\cdot)$, which can retrieve the relevant document $d$ from $\mathbb{D}$, as well as the fine-grained information $u$ from $d$ given query $q$:
\begin{align}
    f(q, \mathbb{D}) \rightarrow \{d\} \\
    f(q, d) \rightarrow \{u\}
\end{align}
In this work, we explicitly define the process as two tasks, \textbf{(1)} the global document retrieval and \textbf{(2)} the local information retrieval, as shown in Figure~\ref{fig:\our}. 

\subsection{Data construction} 
In this work, we consider two main retrieval scenarios: Question Answer Retrieval (QAR) and Relevant Information Retrieval (RIR). In the following sections, we introduce how the data is constructed and outline the specific goals of the retrieval tasks in each scenario.

\noindent\textbf{Question Answer Retrieval} \quad
In this scenario, the query $q$ is in the form of a question, and the goal is to retrieve (1) the reference documents $d$ that support answering the question and (2) the fine-grained sentences $u$ that contain the answer. 



\begin{figure*}[ht]
\begin{center}
  \adjustbox{margin=-0.1cm 0cm 0cm 0cm}{
    \includegraphics[scale=0.55]{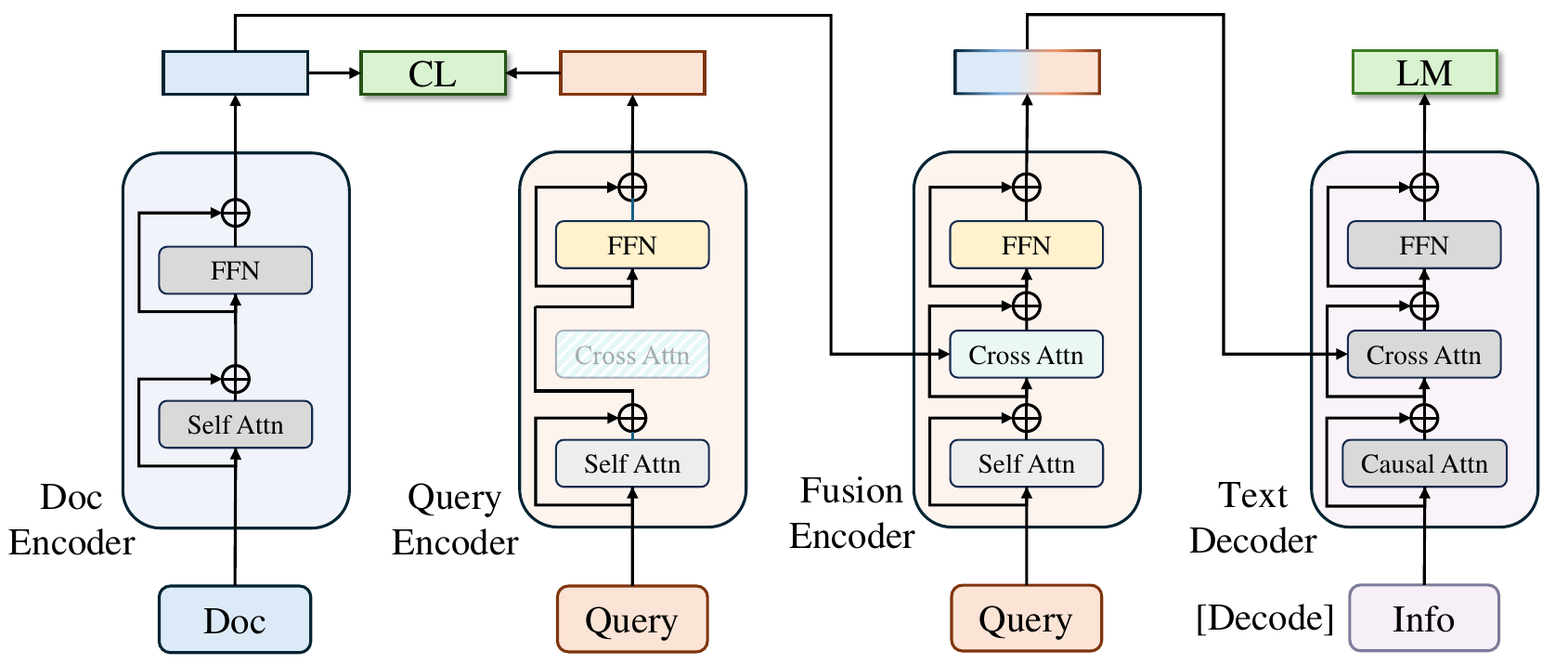}
  }
  \caption{\textbf{\our.} It consists of a bi-encoder, a fusion encoder, and a text decoder. It contains two training objectives, CL represents contrastive learning loss, which aims to optimize the similarity between documents and queries. LM represents the language modeling loss for generating relevant information given documents and queries.}
  \label{fig:model}
\end{center}
\end{figure*}

\noindent\textbf{Relevant Information Retrieval} \quad
This scenario closely mirrors typical user behavior when searching for information on search engines. The query $q$ is typically a few phrases or keywords, the objective is to retrieve (1) the documents $d$ that correspond to the query and (2) the fine-grained sentences $u$ in the documents that are most relevant to the query. However, a significant challenge in this scenario is the difficulty of collecting suitable data from existing public datasets to address this problem. To overcome this, we construct a pipeline to synthesize high quality data using a large language model. Specifically, we select high quality Wikipedia documents~\citep{wikidump}, from which we sampled sentences of appropriate length and whose subject is not a pronoun as $u$. Then we leverage LLM to rewrite these sentences as queries $q$. After applying de-duplication and relevance filtering, we obtain a promising set of \textbf{5.8M} triples. Kindly refer to Appendix~\ref{data} for details on complete data processing procedure.

\subsection{Model Structure}
This section introduces the architecture of \our. It is our intention to enable the model to have both global and local text retrieval capabilities. Inspired by advances in multimodal representation learning~\citep{albef, li2022blip, moco}, we revisit the task from the perspective of modality alignment. Documents and queries can be regarded as two modalities. We facilitate semantic alignment between documents and queries via a bi-encoder, and enable the model to learn to focus on fine-grained query-related information in documents via a fusion encoder and a generation task. The overview of the \our~structure is illustrated in Figure~\ref{fig:model}.

\noindent\textbf{Bi-Encoder}\quad
In the same setup as the classical retrieval approach, we initialize two encoders $E_{d}(\cdot)$ for documents and $E_{q}(\cdot)$ for queries. We use mean pooling to obtain the text embedding.

\noindent\textbf{Fusion Encoder}\quad
The fusion encoder share most of the parameters with query encoder, but have an lightweight learnable cross attention module. In this part, the document embeddings from $E_{d}(\cdot)$ are fused with the query embeddings through cross attention at each layer of the fusion encoder. 

\noindent\textbf{Text Decoder}\quad
The text decoder receives the fusion embeddings and generates fine-grained information\footnote{
Note that in the QAR scenario, the ground truth for the generation is the answer itself, not the full sentence $u$ in which answer appears.} in the document based on the given query and document. It uses a unidirectional causal attention instead of a bidirectional self-attention. A specific 
[Decode]
token is added to identify the beginning of the sequence. The subsequent auto-regressive decoding process will interact with the generated tokens and fusion embeddings to generate text.

\subsection{Training Objectives}
In this section, we introduce the training objectives of \our. 
Through the joint modeling of natural language understanding and natural language generation, \our~can handle global document retrieval and local information retrieval simultaneously.

\noindent\textbf{Contrastive Learning Loss}~(CL)\quad
We use bi-encoder to encode the queries and documents, and optimize the semantic similarity between them through contrastive learning loss~(CL). In addition, we followed the practice in MoCo~\citep{moco} and BLIP~\citep{li2022blip}, where a momentum Bi-Encoder is introduced to encode momentum embeddings and provide richer supervised signals as soft labels.


\noindent\textbf{Language Modeling Loss}~(LM) \quad The introduction of LM loss is crucial for enhancing the local information retrieval capability of \our. LM activates the text decoder, enabling the model to generate relevant intrinsic information by leveraging the fusion embeddings of document and query. It guides the model to learn the fine-grained semantic fusion between query and document. LM optimizes the cross-entropy loss over the entire vocabulary, maximizing the likelihood of the ground truth text. 
The overall loss of \our~is the sum of $\mathcal{L}_\mathrm{CL}$ and $\mathcal{L}_\mathrm{LM}$ with a optional weight $\alpha$:
\begin{align}
\mathcal{L}_\mathrm{\our}=\mathcal{L}_\mathrm{CL}+ \alpha * \mathcal{L}_\mathrm{LM}
\end{align}

\subsection{Inference}\label{infer}
\our's inference process is flexible. In this section, we introduce various usages of \our~to accomplish different tasks.

\noindent\textbf{Global Documents Retrieval}\quad
For this task, we can use the bi-encoder part of \our~to compute the similarity between query and document like the previous classic retrieval method, without introducing any additional parameters and computation cost.

\noindent\textbf{Local Information Retrieval}\quad
The fusion encoder in \our~interacts query and document via cross attention. 
The cross attention weights between each sentence in the document and the query reflect which information the model prioritizes. We rank the sentences based on these weights to retrieve the most relevant fine-grained information from the document.



\section{Experiments}
In this section, we first outline the experimental setup, and then we discuss the overall performance of each task and a more detailed analysis.

\subsection{Setup}

\textbf{Datasets}\quad
For Question Answer Retrieval, we sampled 30M data from PAQ~\citep{lewis-etal-2021-paq} datasets to train \our, and sampled 1M documents and 20k queries as the test set. To verify the generalization ability of methods, we also evaluate the performance on three additional held-out datasets: SQuAD~\citep{rajpurkar-etal-2016-squad}, NQ~\citep{kwiatkowski-etal-2019-nq}, and TriviaQA~\citep{joshi-etal-2017-triviaqa}. For Relevant Information Retrieval, we leverage the synthesized 5.8M data, of which 95\% is used for training and 5\% is reserved for the test set. Specific dataset statistics are in Appendix~\ref{sec:datasets}.

\noindent\textbf{Training Details}\quad
"bert-base-uncased"~\citep{devlin-etal-2019-bert} is used to initialize the encoders in \our. The decoder also has 110M parameters, but is randomly initialized. We train \our~for 10 epochs using batch size of 48 (QAR) / 16 (RIR) on 16 AMD MI200 GPUs. We set the weight $\alpha = 0.25$. We use the AdamW~\citep{adamw} optimizer with a weight decay of 0.05. The full hyperparameters and training settings are detailed in Appendix~\ref{hyper}.






\noindent\textbf{Baselines}\quad
We compare \our~with two types of baselines, one is the text embedding models that have been adequately pre-trained on a large corpus, including SBERT~\citep{sbert}, 
E5~\citep{e5}, BGE~\citep{bge}, GTE~\citep{gte} and ColBERT-QA~\citep{colbert-qa}. The models involved in the comparison are all base versions.
Since the training data of the pre-trained model partially overlaps with the evaluation data, their performance are used as an important reference. 
To ensure a fairer comparison, we retrain SBERT\footnote{\url{https://huggingface.co/sentence-transformers/all-mpnet-base-v2}}~\citep{sbert} and BGE\footnote{\url{https://github.com/FlagOpen/FlagEmbedding}}~\citep{bge} using the open sourced training pipelines with aligned training data and initialization, referred to as SBERT$_{RT}$ and BGE$_{RT}$ in the following. In addition, we also compare with the more complex BGE reranker Base and Large~\citep{bge} in the Local Information Retrieval task.
In the section~\ref{op}, we underline the best performance of the pre-trained models and bold the best performance of the retrained models.



    


\subsection{Overall performance}\label{op}
In this section, we present the overall performance on global document retrieval and local information retrieval.

\noindent\textbf{Global Documents Retrieval}\quad
Firstly, Table \ref{tab:retrieval_pef} reports the comparison with existing methods on global documents retrieval task. We find that \our~delivers competitive performance across multiple datasets even with only tens of millions of training data, demonstrating efficient data utilization. As a reference, the pre-trained SBERT model used 1.17B sentence pairs.
GeAR achieves the state-of-the-art performance on the three datasets SQuAD, PAQ, and RIR, and is slightly weaker than the pre-trained GTE on the NQ dataset. It only lags significantly behind on TriviaQA, but is also better than ColBERT-QA and E5. Compared with ColBERT, GeAR introduces a generation task to explicitly model the alignment relationship between queries and fine-grained semantic fragments of documents, which not only improves the retrieval performance but also reduces the delayed interaction and the increase in space complexity caused by storing multiple vectors. At the same time, GeAR outperforms the retrained model in all metrics. Compared with BGE$_{RT}$, GeAR achieves a relative improvement of 3.5\% in average Recall@5, highlighting the effectiveness of our training method. In Section \ref{4.3}, we further discuss the role of the generation task and its effect on model performance.


\begin{table*}[!ht]
\centering

\newcolumntype{L}{>{\raggedright\arraybackslash}p{2.7cm}} 
\newcolumntype{C}{>{\centering\arraybackslash}p{0.9cm}}  
\hspace*{-6pt}
\begin{tabular}{LCCCCCCCCCC}
\toprule
\multirow{2}{*}{Method} & \multicolumn{2}{c}{SQuAD} & \multicolumn{2}{c}{NQ} & \multicolumn{2}{c}{TriviaQA} & \multicolumn{2}{c}{PAQ} & \multicolumn{2}{c}{RIR}\\
    \cmidrule(lr){2-3} \cmidrule(lr){4-5} \cmidrule(lr){6-7} \cmidrule(lr){8-9} \cmidrule(lr){10-11}
    &R@5 &M@5  &R@5 &M@5  &R@5 &M@5 &R@5 &M@5 &R@5 &M@5 \\
\midrule
\multicolumn{11}{c}{ \quad \textit{Pre-trained retrieval model}} \\
\midrule
SBERT &0.812 &0.667 &0.754 &0.576 &0.677 &0.413 &0.808 &0.701 &0.376 &0.297 \\
E5 &0.803 &0.674 &0.760 &0.581 &0.645 &0.390 &0.816 &0.716 &0.484 &0.396 \\
BGE &0.829 &0.701 &0.674 &0.502 &0.690 &0.422 &0.752 &0.647 &0.451 &0.367 \\
GTE &0.866 & 0.744 &\underline{0.767} &\underline{0.587} &\underline{0.726} &\underline{0.443} &\underline{0.836} &{0.736} &\underline{0.528} &\underline{0.435} \\
ColBERT-QA & \underline{0.882}	& \underline{0.794}	& 0.713	& 0.542	& 0.654 & 0.399 &	0.834 &	\underline{0.755} & - & -\\
\midrule
\multicolumn{11}{c}{ \quad \textit{Retrained retrieval model}} \\
\midrule
SBERT$_{RT}$ &0.742 &0.585 &0.739 &0.550 &0.577 &0.342 &0.859 &0.742 &0.739 &0.631\\
BGE$_{RT}$ &0.841 &0.701 &0.751 &0.553 &0.640 &0.384 &0.901 &0.802 &0.953 &0.881 \\
\midrule
\our
           &0.887 &0.766 &\textbf{0.762} &\textbf{0.574} &\textbf{0.664} &\textbf{0.400} &0.952 &0.872 &\textbf{0.964} &\textbf{0.910} \\
\our$_{w/o \mathcal{L}_\mathrm{LM}}$ 
           &\textbf{0.889} &\textbf{0.776} &0.755 &0.565 &0.660 &0.399 &\textbf{0.955} &\textbf{0.877} &0.963 &0.907 \\
\bottomrule
\end{tabular}
\caption{Comparison of global documents retrieval performance on different datasets, where R@k stands for Recall@k, M@k stands for MAP@k.}
\label{tab:retrieval_pef}

\vspace{5pt} 

\hspace*{-6pt}
\begin{tabular}{LCCCCCCCCCC}
\toprule
\multirow{2}{*}{Method} & \multicolumn{2}{c}{SQuAD} & \multicolumn{2}{c}{NQ} & \multicolumn{2}{c}{TriviaQA} & \multicolumn{2}{c}{PAQ} & \multicolumn{2}{c}{RIR}\\
    \cmidrule(lr){2-3} \cmidrule(lr){4-5} \cmidrule(lr){6-7} \cmidrule(lr){8-9} \cmidrule(lr){10-11}
    &R@1 &M@1  &R@1 &M@1  &R@1 &M@1 &R@1 &M@1 &R@3 &M@3 \\
\midrule
\multicolumn{11}{c}{ \quad \textit{Pre-trained retrieval model}} \\
\midrule
SBERT &0.739 &0.800 &0.558 &0.652 &0.359 &0.583 &0.498 &0.561 &0.891 &0.874 \\
E5 &\underline{0.783} &\underline{0.847} &\underline{0.590} &\underline{0.683} &\underline{0.379} &\underline{0.613} &\underline{0.573} &\underline{0.640} &0.891 &0.878 \\
BGE &0.768 &0.830 &0.570 &0.663 &0.362 &0.589 &0.565 &0.630 &0.894 &0.881 \\
GTE &0.758 &0.820 &0.548 &0.639 &0.352 &0.572 &0.525 &0.590 &\underline{0.895} &\underline{0.886} \\
\midrule
\multicolumn{11}{c}{ \quad \textit{Retrained retrieval model}} \\
\midrule
SBERT$_{RT}$ &0.516 &0.568 &0.445 &0.523 &0.281 &0.472 &0.363 &0.418 &0.899 &0.881 \\
BGE$_{RT}$ &0.455 &0.538 &0.601 &0.656 &0.288 &0.475 &0.409 &0.466 &0.897 &0.888 \\
\midrule
\multicolumn{11}{c}{ \quad 
\textit{Reranker model}} \\
\midrule
BGE-Reranker-B  &0.690 &0.749 &0.641 &0.740 &0.399 &0.640 &0.690 &0.762 &0.884 &0.850\\

BGE-Reranker-L  &0.751 &0.813 &0.670 &0.770 &0.464 &0.737 &0.704 &0.778 &0.891 &0.873\\
\midrule
\our       &\textbf{0.814} &\textbf{0.878} &\textbf{0.761} &\textbf{0.865} &\textbf{0.510} &\textbf{0.797} &\textbf{0.884} &\textbf{0.965} &\textbf{0.933} &\textbf{0.897}\\
\our$_{w/o \mathcal{L}_\mathrm{LM}}$ 
           &0.803 &0.869 &0.582 &0.677 &0.402 &0.650 &0.649 &0.720 &0.891 &0.886 \\
\bottomrule
\end{tabular}
\caption{Comparison of local information retrieval performance on different datasets, where R@k stands for Recall@k, M@k stands for MAP@k.}
\label{tab:localization_pef}
\end{table*}

\begin{figure*}[ht]
  \subfigure[Local information retrieval and generation results of \our~in Question Answer Retrieval scenario.]{\adjustbox{margin=-0.3cm 0cm 0cm 0cm}{
    \includegraphics[scale=0.48]{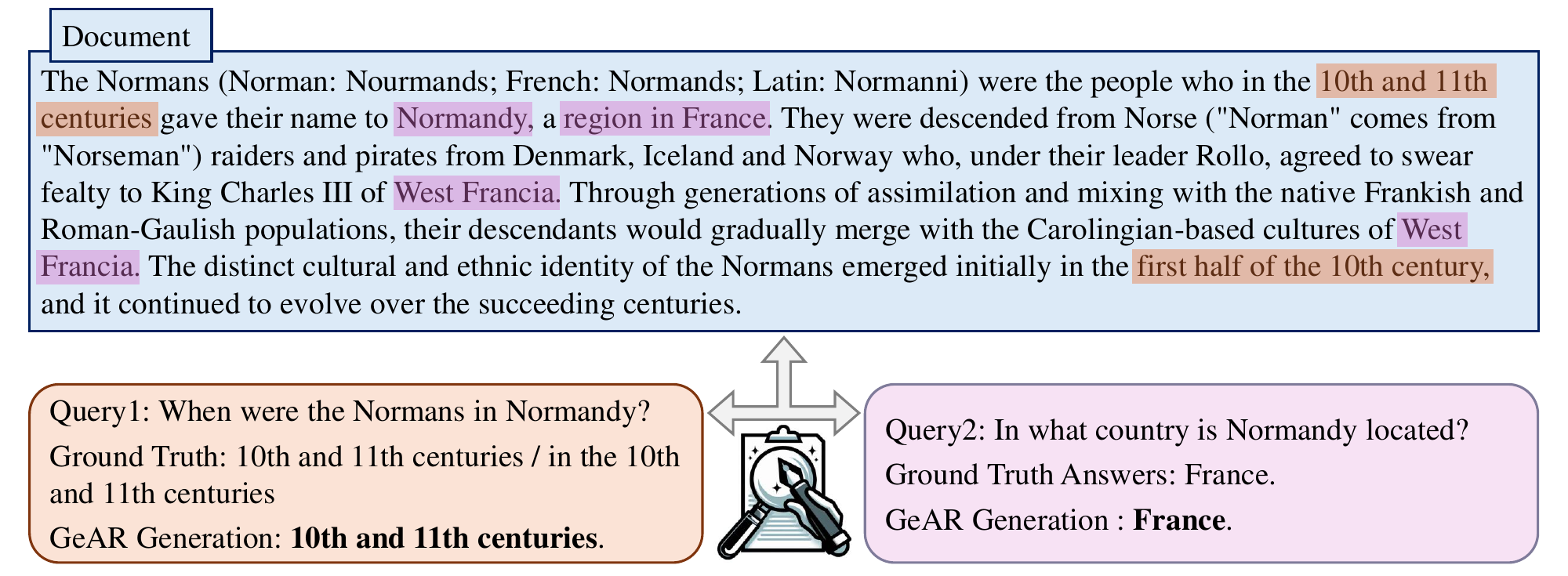}
  }\label{fig:QA_case}}
  \subfigure[Local information retrieval and generation results of \our~in Related Information Retrieval scenario. The sentences in brackets of corresponding colors are the ground truth of the query.]{\adjustbox{margin=-0.3cm 0cm 0cm 0cm}{
    \includegraphics[scale=0.48]{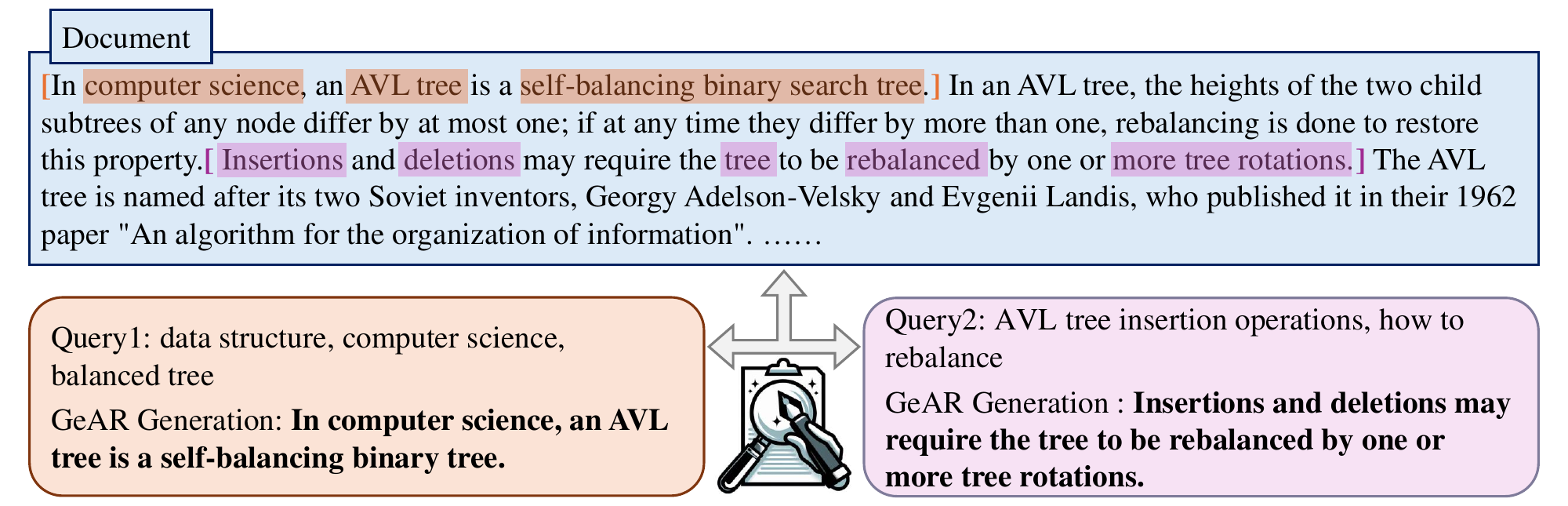}
  }\label{fig:RIR_case}}
  \caption{Visualization of local information retrieval of \our~. In the two scenarios, we pose two different queries for each document and highlight the top 10 tokens with the highest cross attention weights. The tokens with orange background are for \colorbox{orange!30}{query1}, with purple background are for \colorbox{purple!15}{query2}. We also show the generated results of GeAR.}
  \label{fig:Case}
\end{figure*}

\noindent\textbf{Local Information Retrieval}\quad
Next, we evaluate the performance of each method on the local information retrieval task. In the evaluation process, we provide the query and the document $(q, d)$ to the model and observe whether it is able to retrieve the corresponding fine-grained unit $u$. For the retrieval model, we split the documents into sentences and compute their similarity to the query independently, selecting the top-k sentences. In contrast, \our~retrieves units based on the cross attention weights for each sentence given the query, as described in Section \ref{infer}. The results are reported in Table~\ref{tab:localization_pef}. 

It is observed that SBERT$_{RT}$ and BGE$_{RT}$ perform mediocrely, as their training objective focus solely on optimizing the overall similarity between the document and the query, neglecting the fine-grained semantic relationships. The more complex BGE-reranker model performs better than the pure retrieval model. \our~leads the way in all metrics, showing an average relative improvement of 12.9\% over the suboptimal BGE-Reranker-L. Notably, \our~does not require further chunking and encoding of the document. In contrast, \our~benefits from the joint end-to-end training of retrieval and generation, enabling it not only retrieve documents closely aligned with the query but also effectively retrieve fine-grained information within the document.


\begin{table*}[ht]
\centering
\begin{tabular}{lcccccccccc}
\toprule
\multirow{2}{*}{Method} & \multicolumn{2}{c}{SQuAD} & \multicolumn{2}{c}{NQ} & \multicolumn{2}{c}{TriviaQA} & \multicolumn{2}{c}{PAQ} & \multicolumn{2}{c}{RIR} \\
    \cmidrule(lr){2-3} \cmidrule(lr){4-5} \cmidrule(lr){6-7} \cmidrule(lr){8-9} \cmidrule(lr){10-11}
    & EM &F1  &EM &F1  &EM &F1 &EM &F1 &Rouge-1 &Rouge-L \\
\midrule
    Llama 3.2 3B &60.7 &73.3 &57.7 &59.9  &50.4 &66.7 &62.7 &75.5 &69.4 &67.9\\
    Llama 3.3 70B &\textbf{66.2} &\textbf{77.7} &61.0 &\textbf{66.9} &\textbf{56.6} &\textbf{73.0} &61.0 &74.5 &84.4 &84.0\\
    GeAR& 60.0  &64.5 &\textbf{65.7}  &60.7 &46.5  &59.1 &\textbf{87.5}  &\textbf{91.9} &\textbf{87.6} &\textbf{87.3} \\
\bottomrule
\end{tabular}
\caption{Generation performance on different datasets.}
\label{tab:generation}
\end{table*}

\subsection{Analysis}\label{4.3}

\begin{table}[ht]
\centering
\begin{tabular}{ccc}
\toprule
\multirow{2}{*}{$\alpha$} & Global Retrieval& Local Retrieval\\ 
\cmidrule(lr){2-2}  \cmidrule(lr){3-3}
& Ave Recall & Ave Recall\\
\midrule
0 & 0.844 & 0.663 \\ 
0.25 & \textbf{0.846} & 0.781 \\ 
0.5 & 0.844 & \textbf{0.785} \\ 
0.75 & 0.839 & 0.784 \\ 
1 & 0.838 & 0.784 \\ 
\bottomrule
\end{tabular}
\caption{Comparison of performance on two retrieval tasks when the LM loss weight $\alpha$ is varied.}
\label{tune_weight}
\end{table}

\textbf{The Effect of Language Modeling Objectives}\label{ablation}\quad
In this work, we not only optimize the retrieval performance through contrastive learning, but also enhance \our~through the information generation task of a given query, so that it has fine-grained semantic understanding and content retrieval capabilities. We find that if LM loss is removed, both global and local retrieval performance of the model is reduced, as shown in the last row of Table~\ref{tab:retrieval_pef} and Table~\ref{tab:localization_pef}.
Further, we also explore the impact of the weight of LM loss on the overall performance. In Table~\ref{tune_weight}, we observed that the effect of the generation on the retrieval performance is inverted U-shaped, with the optimal values at 0.25 and 0.5 respectively. Higher weights may cause the model to focus on learning the generation task instead, which is similar to previous findings~\citep{mtl}.

\noindent\textbf{Visualization of Local Information Retrieval}\quad
The key distinction between GeAR and traditional retriever is its ability to mine the local information within the document that is most relevant to the query. Figure~\ref{fig:Case} illustrates this process and the generation results of \our~across different scenarios. For each document, we provide two distinct queries and highlight the top 10 tokens with the highest cross attnetion weights corresponding to each query. 
In Figure~\ref{fig:QA_case}, the two queries are related to time and location respectively. \our~not only provides the correct answers but also dynamically adjusts its query-specific focus: it assigns higher attention weights to time-related tokens to the first query and prioritizes tokens related to countries and regions to the second query. 
In Figure~\ref{fig:RIR_case}, depending on the query, \our~focuses on the concept of AVL tree, as well as operations such as insertion and rebalancing, generating corresponding sentences. It is evident that the added generation task enhances the accuracy of local information retrieval. Furthermore, GeAR can not only retrieve detailed content related to the query but also generates corresponding text for reference. This advancement shifts the retrieval results from being mere numerical values to more intuitive and explainable.




\begin{figure}[t]
\begin{center}
  \adjustbox{margin=0cm 0cm 0cm 0cm}{
    \includegraphics[scale=0.62]{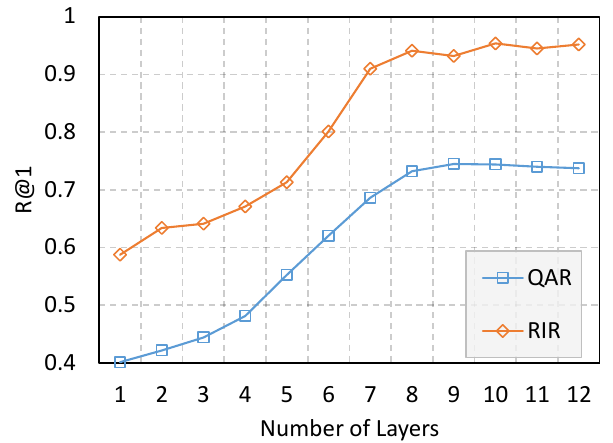}
  }
  \caption{Local information retrieval performance of different layers.}
  \label{fig:loc_with_layer}
\end{center}
\end{figure}

\noindent\textbf{Local Retrieval Performance of Different Layers}\quad
In \our, the query and document tokens interact through the cross attention module at each layer of the fusion encoder. In Figure~\ref{fig:loc_with_layer}, we plot the local retrieval performance using cross attention weights across different layers to examine its relationship with model depth. The results indicate that higher layers generally perform well, as the token embeddings at these layers capture rich semantic information. Interestingly, we observe that the highest layer does not yield the best performance. Instead, peak performance is reached in the last 3 to 4 layers\footnote{In this work, we utilized the 10th layer.}. This phenomenon may arise due to the representations in the highest layer are optimized to serve the final task rather than intermediate interactions. Similar observations have been reported in previous studies involving encoder-only and decoder-only models~\cite{bert-learn, Layers_in_llms}.

\noindent\textbf{Information Generation}\quad
Although generation serves as an auxiliary task in \our~and the decoder is lightweight, we are nonetheless interested in its generation performance. Table \ref{tab:generation} reports the Exact Match (EM) and F1 scores on the QA datasets, and the Rouge~\cite{lin-2004-rouge} scores on the RIR dataset. For reference, we include results from the Llama series model~\citep{llama3}.
Notably, \our~achieves surprising performance on the in-domain data, and performs reasonably well on other test sets. Additionally, Figure~\ref{fig:Case} illustrates examples of \our's ability to generate answers and relevant information, showcasing its satisfactory generation capabilities.




\section{Conclusion}
In this work, to address the challenges of unexplainable and coarse-grained results inherent in current bi-encoder retrieval methods, we propose a direct and effective modeling method: \textbf{Ge}neration \textbf{A}ugmented \textbf{R}etrieval~(\our). 
\our~enhances fine-grained information retrieval by introducing a generation task and incorporating a lightweight decoder and cross attention module, while maintaining the efficiency of the bi-encoder. Experimental results across multiple retrieval tasks and two different scenarios demonstrate that \our~achieves excellent performance and have both global and local understanding and retrieval capabilities. Qualitative analysis further highlights its intuitive and explainable retrieval results. These capabilities make \our~particularly promising in downstream tasks such as web search and retrieval-augmented generation (RAG). We hope that this work offers valuable insights into the gradual unification of natural language understanding and generation paradigms, paving the way for more general and explainable retrieval systems in the future.




\section*{Limitations}

Due to constraints in computational resources and associated costs, the synthesized data used in our experiments is not as comprehensive as that found in traditional retrieval scenarios. While the results demonstrate the efficacy of \our, applying it to more diverse and semantically rich retrieval scenarios remains an important direction for future exploration.


Additionally, the context length of \our~is limited to 512 tokens, consistent with the chunk lengths commonly used in retrieval tasks. However, recent advancements in extending the context length of retrieval models, such as those proposed in~\citep{longembed}, suggest exciting opportunities to overcome this limitation. Extending \our's context length could further enhance its capabilities in handling long-form retrieval tasks, which we plan to investigate in future work.

Thirdly, the decoder of GeAR has only 110M parameters, the same as the encoder. Moreover, the focus of GeAR is not to optimize the generation performance of the model, and the generation task is not the main task. Therefore, GeAR cannot complete other complex generation tasks like Llama~\citep{llama3}. In future work, whether GeAR can be scaled up to enable it to complete retrieval tasks and respond well to various generation problems will be an interesting direction.


We hope that the above discussions can inspire further investigation within the research community, encouraging advancements that address these limitations and contribute to the broader progress of NLP research.

\bibliography{anthology,custom}
\bibliographystyle{acl_natbib}

\appendix

\section*{Appendices}\label{sec:appendix}

\section{Data Construction}\label{data}
We present here the practice of synthesizing data for Relevant Information Retrieval scenarios.

\textbf{Pre-processing}\quad
Firstly, we choose high-quality documents from Wikipedia~\citep{wikidump}. We process the documents sentence by sentence, removing sentences with repetitive line breaks and phrases, until the document processing is complete or the token count reaches 500 (<512). We remove the documents that are too short, with a sentence count less than 3 or a token count of less than 200. Second, we filter the candidate sentences in the document that can be rewritten: we filter all the sentences that have a token count between 8 and 20 and whose first word and subject are not pronouns (the set of pronouns includes {"this", "these", "it", "that", "those", "they", "he", "she", "we", "you", "I"}). If the number of sentences filtered is less than 3, we discard the document.

\textbf{LLM Rewriting}\quad 
We randomly select 3 sentences in the document and use vLLM~\citep{vllm} and "Llama-3.1-70B-Instruct"~\citep{llama3} to rewrite them into queries, the prompt is: "You are a helpful assistant, please help the user to complete the following tasks directly, and answer briefly and fluently. This is a sentence from Wikipedia. Assuming that users want to search for this sentence on a search engine, write a phrase that users might use to search (including some keywords), separated by commas. Retain the key information of the subject, object, and noun. Unimportant words can be modified, but do not add other information.".

\textbf{Post-processing}\quad
We de-duplicate the keywords in the rewritten query and then reorder them. To ensure the relevance of the query to the document, we perform a round of filtering using BGE~\citep{bge} to retain the data with a similarity of 0.5 or more between the rewritten query and the document. In this way we obtain a reasonable triad of queries, documents, and units (sentences).

For the construction of Relevant Information Retrieval data, we have also tried to collect paired sentences and make LLM expand one of them into a document. However, we fine that other sentences in the LLM expansion were less informative than the original sentence, for example, being some descriptive statements were generated around the original sentence. This pattern tends to cause the model to learn to locate the central sentence, or the most informative sentence, in the expanded document, leading model to ignore the query. So please be aware of this if you plan to try this way of constructing your data.

\begin{table}[!ht]
\centering
\resizebox{\columnwidth}{!}{%
\begin{tabular}{ll}
\bottomrule  
\textbf{Hyperparameter} & \textbf{Assignment}        \\ \bottomrule
Computing Infrastructure & 16 MI200-64GB GPUs         \\
Number of epochs        & 10                          \\ 
Batch size per GPU      & 48 / 16                     \\ 
Maximum sequence length & 512                       \\ 
Optimizer               & AdamW                       \\ 
AdamW epsilon            & 1e-8                       \\ 
AdamW beta weights       & 0.9, 0.999                  \\ 
Learning rate scheduler & Cosine lr schedule              \\ 
Initialization learning rate   & 1e-5 \\ 
Minimum learning rate   & 1e-6 \\ 
Weight decay            & 0.05                       \\ 
Warmup steps            & 1000                        \\ 
Warmup learning rate & 1e-6         \\
\bottomrule
\end{tabular}%
}
\caption{Hyperparameter settings}
\label{tab: training parameters}
\end{table}

\section{Overview of datasets}\label{sec:datasets}
We describe here in detail the datasets used for training and evaluation. 

\subsection{Training}
For Question Answer Retrieval, we sampled 30M data from PAQ~\citep{lewis-etal-2021-paq} datasets to train \our. For Relevant Information Retrieval, we used the 95\% of the synthetic data for training. The specific statistics are shown in Table \ref{tab:train_data}.

\begin{table}[h]
    \centering
    \begin{tabular}{cc}
    \toprule
        Scenario & Data Number \\
    \midrule
        QAR & 30,000,000 \\
        RIR & 5,676,877 \\
    \bottomrule
    \end{tabular}
    \caption{Training data statistics.}
    \label{tab:train_data}
\end{table}

\begin{table*}[t]
\centering
\begin{tabular}{llcc}
\toprule
\textbf{Scenario} & \textbf{Dataset} & \textbf{Documents Number} & \textbf{Queries Number} \\
\midrule
\multirow{4}{*}{QA} & Squad    & 20,239  & 5,928  \\
                    & NQ       & 64,501  & 2,889  \\
                    & TriviaQA & 104,160 & 14,000 \\
                    & PAQ      & 932,601 & 20,000 \\
\midrule
RIR                 & RIR      & 2,315,413 & 145,562 \\
\bottomrule
\end{tabular}
\caption{The evaluation data statistics for the global document retrieval task.}
\label{tab:data_for_retrieval}
\end{table*}

\begin{table*}[t]
\centering
\begin{tabular}{llc}
\toprule
\textbf{Scenario} & \textbf{Dataset} & \textbf{Data Number} \\
\midrule
\multirow{4}{*}{QA} & Squad    & 5,928  \\
                    & NQ       & 2,889  \\
                    & TriviaQA & 14,000 \\
                    & PAQ      & 20,000 \\
\midrule
RIR                 & RIR      & 10,000 \\
\bottomrule
\end{tabular}
\caption{The evaluation data statistics for the local information retrieval and generation tasks.}
\label{tab:data_for_loc}
\end{table*}

\subsection{Evaluation}
In the evaluation stage, we introduce the specific information of the evaluation data by task.

\textbf{Global Documents Retrieval}\quad
First, for the global document retrieval task, the queries come from the test set in the respective dataset, and the candidate documents are all documents within the entirety of the dataset, including the SQuAD~\citep{rajpurkar-etal-2016-squad}, NQ~\citep{kwiatkowski-etal-2019-nq}, TriviaQA~\citep{joshi-etal-2017-triviaqa}, and RIR datasets. 
It is difficult to encode all the documents of the PAQ dataset because the dataset is too large. So for the PAQ dataset, we sampled 1M documents and 20k queries, all of which have no intersection with the training data. The evaluation data statistics for the document retrieval task are shown in Table \ref{tab:data_for_retrieval}.

\textbf{Local Information Retrieval and Generation} \quad
For these two tasks, we directly use the test set data corresponding to the respective datasets. Therefore, their number is consistent with the number of queries in Table \ref{tab:data_for_retrieval}. For the RIR dataset, we sample 10k records as the test set. The evaluation data statistics for the local information retrieval and generation tasks are shown in Table \ref{tab:data_for_loc}.

\section{HyperParameters and Implementation Details}\label{hyper}
We run model training on 16 AMD MI200 GPUs with 64GB memory and evaluation on 8 NVIDIA Tesla V100 GPUs with 32GB memory. The learning rate is warmed-up from 1$e$-6 to 1$e$-5 in the first 1000 steps, and then following a cosine scheduler, where the mininum learning rate is 1$e$-6. The momentum parameter for updating momentum encoder is set as 0.995, the queue size is set as 57600. We linearly ramp-up the soft labels weight from 0 to 0.4 within the first 2 epoch. The overall hyperparameters are detailed in Table~\ref{tab: training parameters}. We use FAISS \citep{faiss, faiss-gpu} to store and search for vectors. The 2 encoders and 1 decoder in \our~are the same size as "bert-base"~\cite{devlin-etal-2019-bert}, the total number of parameters of \our~is about 330M. The training time for QAR scenario is about 5 days, for RIR scenario is about 3 days. 

\begin{table*}[t]
\centering
\begin{tabular}{lcccc}
\toprule
 & CPU & Ratio (vs BGE) & GPU & Ratio (vs BGE) \\
\midrule
BGE & 225s & 1x & 34s (1 * H20) & 1x \\
GeAR & 288s & 1.28x & 56s (1 * H20) & 1.65x \\
BGE Reranker-large & - & - & 285s (8 * H20) & 8.38x (8 * H20) \\
\bottomrule
\end{tabular}
\caption{Comparison of the inference cost of the models on the local information retrieval task.}
\label{tab:inference_cost}
\end{table*}

\section{Inference Cost}
In our method, GeAR implements three distinct forward processes:

\textbf{Global Retrieval}: Uses only the bi-encoders for global retrieval, with computational complexity identical to classical retrieval models. Document embeddings can be precomputed offline.

\textbf{Local Retrieval}: Computes fusion encoder cross-attention weights for local retrieval without decoder involvement.

\textbf{Generation}: Activates the decoder only when needed, to generate text.

In the local retrieval, GeAR may introduce some inference cost compared to the classic bi-encoder. Therefore, we tested the inference cost on 2889 test data of NQ task in a CPU environment (AMD EPYC 9K84 96-Core Processor * 2) and a GPU (NVIDIA H20) environment, the results are reported in Table~\ref{tab:inference_cost}.

We observe that GeAR consumes about 1.28x (on CPU) and 1.65x (on GPU) of the same-sized bi-encoder, which is a moderate constant time increase. We also tested the inference speed of BGE Reranker-large on 8 * H20. Since BGE reranker is a complex cross encoder, it needs to fully interact and score each query and candidate. Therefore, its time consumption on 8 * H20 is still 8.38x that of BGE.

\section{More Visualization}
\begin{figure*}[ht]
\begin{center}
  \adjustbox{margin=-0.25cm 0cm 0cm 0cm}{
    \includegraphics[scale=0.48]{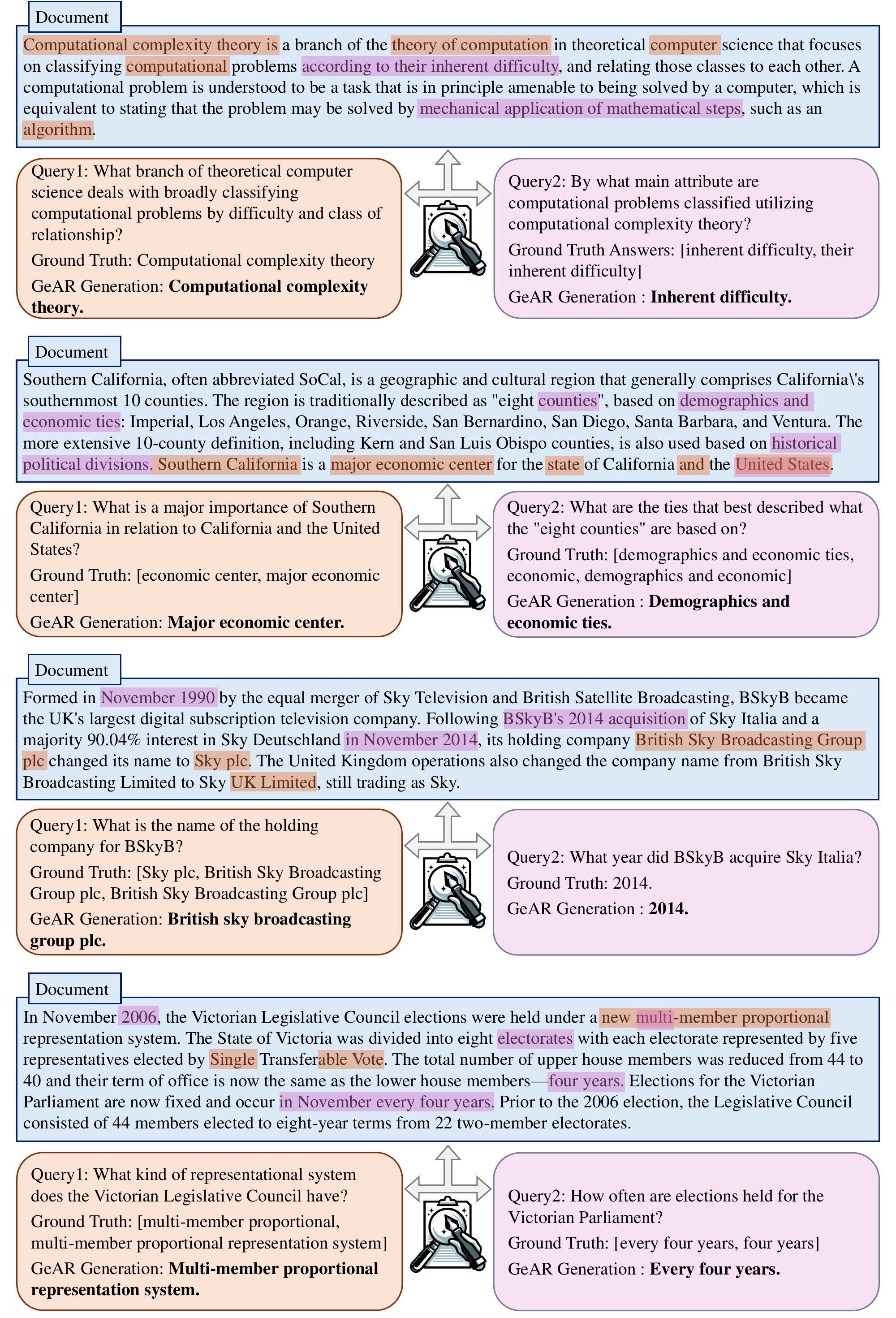}
  }
\end{center}
\caption{More Visulization results.}
\label{fig:more_case}
\end{figure*}

To present the effect of \our~intuitively, we show more visualisation results of \our~in Figure~\ref{fig:more_case}. Each example contains two different queries for a document to observe whether GeAR can respond differently to different queries, including locating key information and generating answers. We also highlight the top 10 tokens with the highest cross attention weights for the corresponding queries. The tokens with orange background are for \colorbox{orange!30}{query1}, and the tokens with purple background are for \colorbox{purple!15}{query2}. 

\end{document}